\documentclass[letterpaper,12pt]{article}
\usepackage[hmargin=1in,vmargin=1in]{geometry}

\usepackage{graphicx} 

\usepackage{float}
\usepackage{multirow}

\usepackage{booktabs}
\usepackage{graphicx}

\newcounter{remark}

\newenvironment{Remark}{
\vspace{10pt}
\refstepcounter{remark}
\noindent{\bf Remark \theremark.}\
}{
\par }

\usepackage{listings}
\usepackage{amsmath,amsfonts,amssymb,amsthm,epsfig,xspace,graphics,graphicx,mathtools}

\usepackage{setspace}
\usepackage{framed}
\usepackage{hyperref}
\usepackage{cancel}

\usepackage[round,authoryear]{natbib}
\usepackage[normalem]{ulem}

\graphicspath{{./Figures/}}

\newcommand{\trans}{{\mskip-2mu\scriptscriptstyle\top}} 

\usepackage{pdfpages} 

\newcommand{\Def}{\overset{\text{def}}{=}}

\newcommand{\Bt}{\calB_t}  
\newcommand{\Br}{\calB_\mat}  
\newcommand{\St}{\calS_{\breve{\bft}}}

\newcommand{\Su}{\calS_{\breve{\bfu}}}  

\newcommand{\Sdnonlocal}{\calS_{\nu_s^{\text{nonlocal}}}}

\newcommand{\Sgradnunonlocal}{\calS_{\grad \nu_s^{\text{nonlocal}}}}

\newcommand{\zed}{{\bf 0}}
\newcommand{\id}{{\bf 1}}

\newcommand{\vvs}{\mskip1mu}
\newcommand{\mat}{\text{\tiny R}}%

\newcommand{\grad}{\text{grad}\,}
\newcommand{\divx}{\text{div}\vvs}

\newcommand{\tendot}{\mskip-3mu:\mskip-2mu}

\newcommand{\calB}{\mathcal{B}}%
\newcommand{\calE}{\mathcal{E}}%
\newcommand{\calL}{\mathcal{L}}%
\newcommand{\calS}{\mathcal{S}}%
\newcommand{\bfA}{{\bf A}}%
\newcommand{\bfb}{{\bf b}}\newcommand{\bfB}{{\bf B}}%
\newcommand{\bfC}{{\bf C}}%
\newcommand{\bfF}{{\bf F}}%
\newcommand{\bfI}{{\bf I}}%
\newcommand{\bfK}{{\bf K}}%
\newcommand{\bfn}{{\bf n}}%
\newcommand{\bfR}{{\bf R}}%
\newcommand{\bft}{{\bf t}}\newcommand{\bfT}{{\bf T}}%
\newcommand{\bfu}{{\bf u}}\newcommand{\bfU}{{\bf U}}%
\newcommand{\bfx}{{\bf x}}%
\newcommand{\bfchi}{\boldsymbol{\chi}}%

\newcommand{\dis}{\text{dis}}
\newcommand{\vol}{\text{vol}}

\newcommand{\skw}{\hbox{\rm skw}\mskip3mu}
\newcommand{\sym}{\hbox{\rm sym}\mskip3mu}
\newcommand{\Tr}{\hbox{\rm tr}\mskip2mu}

\newcommand{\Div}{\hbox{\rm Div}\mskip2mu}

\title{A non-local constitutive model for the Mullins effect in filled elastomers}

\author{Keven Alkhoury${^\dagger}$\thanks{Corresponding author: Keven$\_$Alkhoury@brown.edu; Keven.Alkhoury@gmail.com}\footnotemark[1]\\
\ \\
${^\dagger}$ School of Engineering\\
Brown University\\
184 Hope Street, Providence, RI 02906 USA\\
}

\begin{document}

\maketitle


\begin{abstract}
Filled rubber-like materials are widely used in engineering applications and are well known to exhibit the Mullins effect. In this work, an established local constitutive model from the literature is extended to a non-local setting to resolve the mesh dependence inherent to the local approach. Non-local effects are incorporated using two separate approaches: (i) a Helmholtz-type equation governing a non-local soft volume fraction, and (ii) a Laplacian term introduced directly into the soft volume fraction local evolution law. In both formulations, an additional governing partial differential equation arises and is solved numerically in Abaqus using an analogy with the heat equation. The two approaches yield different results, leaving the choice between them to be guided by experimental findings. The details of the implementation, along with the code developed in this work are also provided.

\end{abstract}


\section{Introduction}
Filled rubber-like materials, also known as filled elastomers, consist of a polymer matrix embedded with stiff filler particles and are used in many applications ranging from industrial and consumer \citep{clark1981mechanics, Leblanc2002, toopchi2008lateral} to medical \citep{birmingham1998effect, herrington2005effect}. In general, filled elastomers exhibit nonlinear inelastic phenomena, such as the Mullins effect.

The Mullins effect was first intensively studied by \citet{Mullins1948} more than 7 decades ago, yet there is still no general agreement on either the physical source or the mechanical modeling of this effect \citep{Diani2009,PLAGGE2019100588,krebs2026mullins}.
While significant progress has been made in understanding the Mullins effect, existing modeling approaches remain largely either phenomenological or micro-mechanically motivated. For a detailed and comprehensive discussion, the interested reader is referred to existing review articles and our earlier work on the subject \citep[cf. eg.,][and references within]{Diani2009,alkhoury2024experiments,krebs2026mullins,nyevgen2026thermal}.

In a recent contribution, \citet{alkhoury2024experiments} characterized and modeled the behavior of a handful of commercially available filled rubber-like materials, building on the framework of \citet{qi2004constitutive}, in which filled elastomers are treated as composite materials consisting of a soft polymer matrix filled with stiff filler particles.
In their work, the filler volume fraction is denoted by $\nu_f$, while the volume fraction of the soft polymeric domain is given by $\nu_s = 1 - \nu_f$, and the material is assumed to consist primarily of hard regions that progressively transform into soft regions through an evolution equation governing the soft volume fraction \citep{alkhoury2024experiments}. Accordingly, the soft volume fraction $\nu_s$ is treated as a local internal variable, as in \citet{qi2004constitutive}; as a result, its evolution is solved at the material point level, leading to mesh dependence inherent to the local formulation.

The objective of this work is to overcome this challenge by incorporating non-local effects into the evolution of the soft volume fraction $\nu_s$ through two approaches: (i) a Helmholtz-type equation governing a non-local soft volume fraction, and (ii) a Laplacian term introduced directly into the local evolution law. In both formulations, an additional governing partial differential equation arises and is solved numerically in Abaqus using an analogy with the heat equation, following our recent work \citep{ALKHOURY2026}. The resulting formulations are mesh-independent, and yield different results, leaving the choice between them to be guided by experimental findings. 
The novelty of this work lies in providing both local and non-local formulations of the Mullins effect within a unified framework, together with its numerical implementation and corresponding code using Abaqus subroutines. To the best of the author’s knowledge, no existing work provides such a combined formulation and implementation.

The remainder of this paper is organized as follows. In Section \ref{sec:Continuum}, we overview the constitutive framework, summarize the existing ``local'' constitutive model, and present our ``non-local'' extension. In Section \ref{sec:FE_Implementation}, we explore the similarities between the heat equation and the ``non-local'' soft volume fraction evolution by examining Abaqus documentation. We then detail the implementation of the ``non-local'' constitutive model. In Section \ref{sec:Applications}, we showcase the usefulness of our work by looking into a boundary value problem. We provide concluding remarks in Section \ref{sec:Conclusion}.

\section{Continuum framework}\label{sec:Continuum}

In this section, we provide an overview of the kinematics and the continuum-level governing equations that describe the nonlinear mechanical behavior of elastomers, including the Mullins effect.

\subsection{Kinematics}

Consider an undeformed body $\calB_\mat$ identified with the region of space it occupies in a fixed reference configuration, and denote by $\bfx_\mat$ an arbitrary material point of $\calB_\mat$. The referential body $\calB_\mat$ then undergoes a motion $\bfx = \bfchi(\bfx_\mat,t)$ to the deformed body $\calB_t$ with deformation gradient given by\footnote{Following common notation \citep{gurtin2010mechanics}, the symbols $\nabla$ and $\Div$ denote the gradient and divergence with respect to the material point $\bfx_\mat$ in the reference configuration; while $\grad$ and $\divx$ denote these operators with respect to the point $\bfx = \bfchi(\bfx_\mat,t)$ in the deformed configuration.
Also, we write $\Tr \bfA$, $\sym \bfA$, $\skw \bfA$, and $\bfA_{0}$ respectively, for the trace, symmetric, skew, and deviatoric parts of a tensor $\bfA$. Lastly, the inner product of tensors $\bfA$ and $\bfB$ is denoted by $\bfA \tendot \bfB$, and the magnitude $\bfA$ by $\left| \bfA \right| = \sqrt{\bfA \tendot \bfA}$.} 
\begin{equation}\label{eqn:DefGrad}
  \bfF = \nabla \bfchi, \quad\text{such that}\quad J=\det\bfF>0. 
\end{equation}
The left and right Cauchy-Green deformation tensors are given by
\begin{equation}
\bfB=\bfF\bfF^{\trans} \,,
\end{equation}
and
\begin{equation}
\bfC=\bfF^{\trans}\bfF \,.
\end{equation} 
Also, the polar decomposition of the deformation gradient
\begin{equation}
  \bfF = \bfR\bfU
\end{equation}
allows its split into a rotation $\bfR$, and a symmetric stretch $\bfU$.

Since elastomers are typically \emph{nearly}-incompressible, we introduce the distortional and volumetric parts of the deformation gradient, defined as 
\begin{equation}
  \bfF_\dis = J^{-1/3}\bfF \quad \text{where} \quad \det\bfF_\dis=1 \,, 
\end{equation}
and
\begin{equation}
  \bfF_\vol = J^{1/3}\bfI \quad \text{where} \quad \det\bfF_\vol=J \,, 
\end{equation}
so that
\begin{equation}
  \bfF = \bfF_\dis \bfF_\vol  \,.
\end{equation}
The corresponding distortional left and right Cauchy-Green deformation tensors are then 
\begin{equation}
\bfB_\dis = \bfF_\dis\bfF_\dis^{\trans}=J^{-2/3}\bfB \,,
\end{equation}
and
\begin{equation}
\bfC_\dis = \bfF_\dis^{\trans}\bfF_\dis=J^{-2/3}\bfC \,.
\end{equation} 

\subsection{Stress softening variables}
Building upon literature \citep{mullins1957theoretical, mullins1965stress, harwood1965stress, harwood1966stress}, and following our recent work \citep{alkhoury2024experiments}, we treat filled elastomers as a soft polymer matrix filled with stiff filler particles where the filler volume fraction is denoted by $\nu_f$, and the volume fraction of soft polymeric domain is $\nu_s = 1 - \nu_f$, and the virgin material is assumed to have primarily hard regions that deform to soft regions.

Additionally, according to \citet{mullins1957theoretical}, when filled elastomers undergo an arbitrary deformation, the hard filler accommodates much less of the overall deformation than the soft rubber matrix. Therefore, an amplified stretch
\begin{equation}\label{eq:Lambda}
\Lambda = \sqrt{X (\bar{\lambda}^2-1)+1}
\end{equation}
is used as a deformation measure for the the polymeric matrix, where $X$ is an amplification factor, and $\bar{\lambda}$ is the effective stretch given by
\begin{equation}\label{eq:Lambdabar}
 \bar{\lambda}=\sqrt{\Tr\bfC_\dis/3} \,.
\end{equation}
The amplification factor $X$ depends on the soft volume fraction $\nu_s$ and the shape of the filler particles.  In this work, we adopt the standard form proposed by \citet{guth1945theory} 
\begin{equation}\label{eq:AmplificationFactor}
X =1 + 3.5 (1-\nu_s) + 18 (1-\nu_s)^2 \,.
\end{equation}

\subsection{Summary of the existing ``local'' constitutive model
}\label{sec:Summary}

\subsubsection{Free energy}\label{sec:FreeEnergy}
We adopt the work of \citet{qi2004constitutive} by assuming that a typical filled elastomer material may be treated as a composite material with rigid filler particles, and take the free energy density per unit reference volume to be  
\begin{multline}
\hat{\psi}_\mat(\bfC,\nu_s)  = 
 \nu_s  G_0 {\lambda_L}^{2}\left[ \left( \frac{\Lambda}{\lambda_L} \right) \beta + \ln\left( \frac{\beta}{\sinh\beta}   \right) - \left( \frac{1}{\lambda_L}\right) \beta_0 - \ln\left(\frac{\beta_0}{\sinh\beta_0}\right)\right] \\
   + \frac{1}{2} K ({\ln J})^{2} \,,
\label{eqn:FreeEnergy}
\end{multline}
where $\nu_s$ is the soft volume fraction, $G_0$ is the initial shear modulus,  $\lambda_L$ is the locking stretch, and $\Lambda$ is the amplified stretch previously described in \eqref{eq:Lambda}. Moreover, $\beta$ and $\beta_0$ are functions given by
\begin{equation} 
\beta = \calL^{-1}\left(\dfrac{\Lambda}{\lambda_L}\right) \quad \text{and} \quad \beta_0 = \calL^{-1}\left(\dfrac{1}{\lambda_L}\right),   
\end{equation}
where  $\calL^{-1}$ is the inverse of the Langevin function, $\calL (\bullet) = \coth (\bullet) - 1/(\bullet)$. $K$ is the bulk modulus used to approximate the \emph{near-incompressible} conditions and is assumed to be three orders of magnitude greater than the shear modulus $G_0$ (i.e., $K=10^3 G_{0}$).

\subsubsection{Cauchy stress}
Based on thermodynamic restrictions, straightforward calculations provide the Cauchy stress $\bfT$ in the form
\begin{equation}
\bfT=J^{-1}\bfF\left(2\dfrac{\partial\hat{\psi}_\mat}{\partial\bfC}\right)
\bfF^{\trans} = J^{-1}\big[ G \left(\bfB_\dis\right)_{0} + K \left(\ln J\right) \id \big]
  \, ,
\label{eq:CauchyStress} 
\end{equation}
where the shear modulus $ G = \nu_s X G_0\left(\frac{\lambda_L}{3\Lambda}\right) \calL^{-1}\left(\frac{\Lambda}{\lambda_L}\right)$ is a function of the amplified stretch and stress softening variables. 

\subsubsection{Evolution equations}
Building upon \citet{qi2004constitutive} and our previous work \citep{alkhoury2024experiments}, we take the evolution of the soft domain fraction to be
\begin{equation}\label{eq:EvolutionSoftDomain}
\dot{\nu_s} = A (\nu_{ss}-\nu_s) \frac{\lambda_{L} - 1}{(\lambda_{L} - \Lambda^\text{max})^2} \dot{\Lambda}^\text{max} \,, \qquad \nu_s (\bfx_\mat, t = 0) = \nu_{s0}\,,
\end{equation}
with
\begin{equation}\label{eq:LamDotMax}
  \dot{\Lambda}^\text{max} =\begin{cases}
    \dot{\Lambda}, &  \Lambda = \Lambda^\text{max}\,,\\
      0 &  \Lambda < \Lambda^\text{max}\,,
  \end{cases} 
\end{equation}
and $A > 0$ a material parameter. Additionally, \eqref{eq:EvolutionSoftDomain} states that as $\nu_s$ approaches the steady state saturation value of $\nu_{ss}$ whenever $\dot{\Lambda} \neq 0$ starting from an initial condition $\nu_{s0}$ in the virgin state.  Moreover, according to \citet{qi2004constitutive}, $\nu_s$ approaches its steady state saturation value $\nu_{ss}$ faster than $\Lambda^\text{max}$ approaches $\lambda_L$, for that reason, $\dot{\nu_s}$ will always become dormant prior to chain locking preventing any associated numerical issues.

\subsubsection{Governing equations}
%

Neglecting inertial effects, the balance of forces and moments in the deformed body $\Bt$ are expressed as 
\begin{equation}\label{eqn:BalanceForcesMoments_Spatial}
  \divx\bfT  + \bfb = \zed\quad\text{and}\quad\bfT = \bfT^{\trans},
\end{equation}
where $\bfT$ and $\bfb$ represent the Cauchy stress provided in \eqref{eq:CauchyStress} and an external body force per unit deformed volume, respectively. Moreover, the standard boundary conditions are prescribed displacement and tractions
\begin{equation}\label{eqn:BalanceForcesMoments_Spatial_BC}   
  \left.
  \begin{aligned}
       \bfu = \breve{\bfu} \qquad  \qquad &\text{on} \quad \Su\,,\\
       \bfT \bfn = \bf{\breve{t}} \qquad  \qquad &\text{on} \quad \St\,,
  \end{aligned}  
  \right\}
\end{equation}
where $\Su$ and $\St$ represent complementary subsurfaces of the boundary $\partial\Bt$ of the body $\Bt$, such that, $\partial\Bt = \Su \cup \St$ and $\Su \cap \St = \varnothing$, with an initial condition $\bfu (\bfx_\mat,0) = \bfu_{0} (\bfx_\mat)$ in $\Br$.

\subsubsection{Model features through material point calculations}
\begin{figure}[H]
\begin{tabular}{cc}
\includegraphics[width = .5\textwidth]{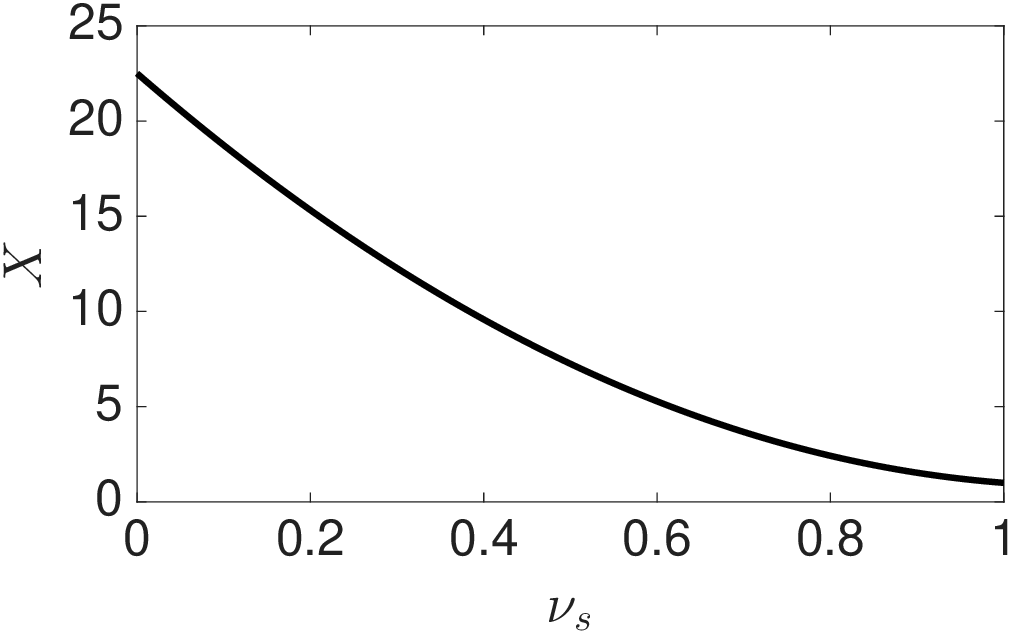}  & \includegraphics[width = .5\textwidth]{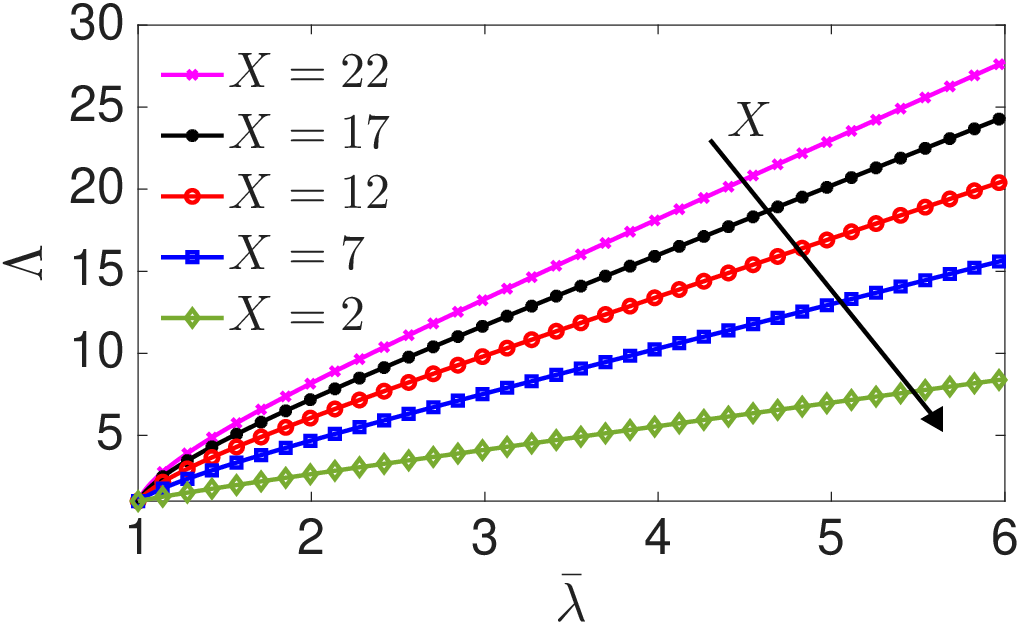}\\
a) & b)
    \end{tabular}
	\caption{Material-point response of the local model of \citet{qi2004constitutive}: a) amplification factor $X$ evolution with soft volume fraction $\nu_s$, and b) amplified stretch $\Lambda$ evolution with effective stretch $\bar{\lambda}$ for various amplification factors $X$.}
\label{fig:MaterialPointCalc}
\end{figure}

In this section, we showcase the ``local'' model features of \citet{qi2004constitutive} through material-point calculations. As noted earlier, it is \emph{idealized} that deformation of the virgin material causes hard regions to transition into soft regions. Since the hard filler accommodates much less of the overall deformation than the soft rubber matrix, an amplification factor $X$ is introduced in \eqref{eq:AmplificationFactor}.
Figure \ref{fig:MaterialPointCalc}a represents the dependence of $X$ on $\nu_s$, indicating that $X$ decreases as the material undergoes the hard-to-soft transition during deformation.
Figure \ref{fig:MaterialPointCalc}b shows the evolution of the amplified stretch $\Lambda$ introduced in \eqref{eq:Lambda} with deformation, represented by the effective stretch $\bar{\lambda}$ introduced in \eqref{eq:Lambdabar}, for various amplification factors $X$. It can be observed that $\Lambda$ increases with deformation, and, for a fixed deformation, attains larger values as $X$ increases. This is an important feature of the model, as a decrease in the hard volume fraction in the deforming material undergoing a hard-to-soft transition requires a smaller $\Lambda$ to accommodate hard-filler deformation, which is known to be much smaller than the overall deformation of the soft rubber matrix.

\begin{figure}[H]
\begin{tabular}{cc}
\includegraphics[width = .5\textwidth]{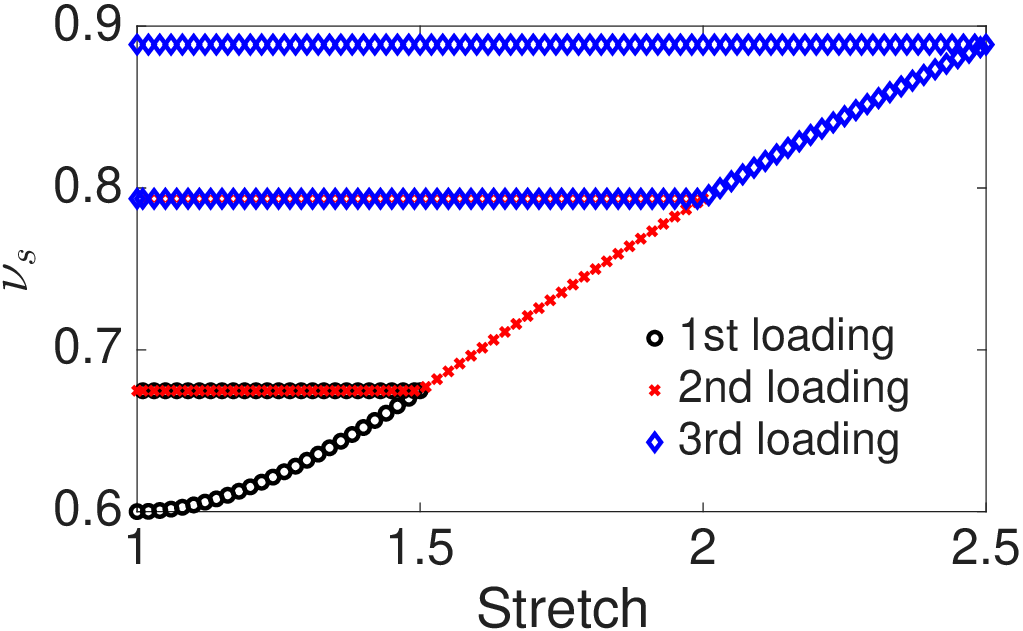}  & \includegraphics[width = .5\textwidth]{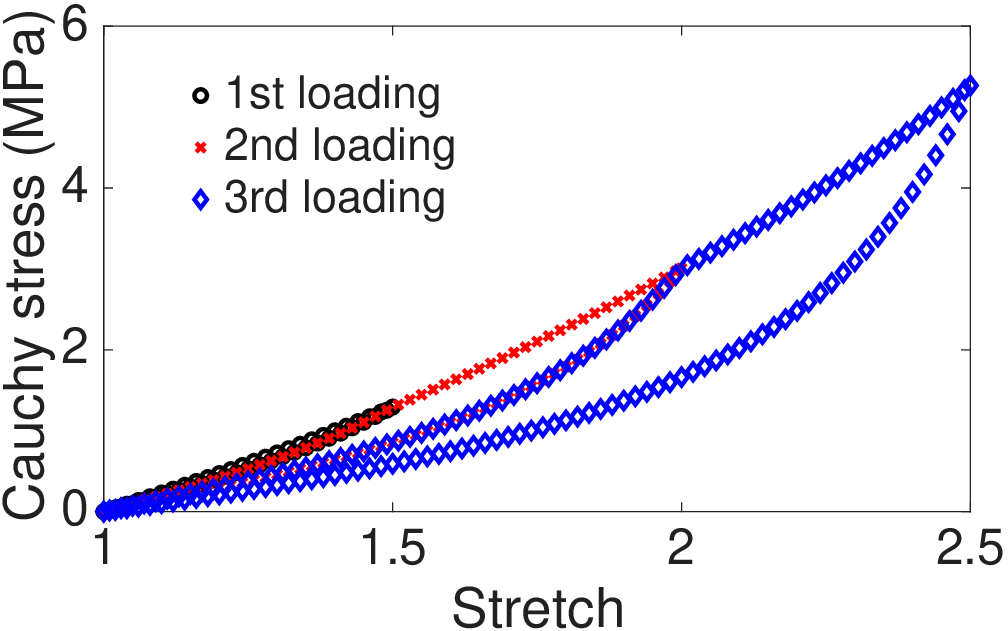}\\
a) & b)
    \end{tabular}
	\caption{Material-point response of the local model of \citet{qi2004constitutive}: a) evolution of the soft volume fraction $\nu_s$ for a load/unload/reload cyclic profile, and b) the corresponding Cauchy stress vs. stretch response, for a representative material.}
\label{fig:StressStretchMullins}
\end{figure}

Moreover, Figure \ref{fig:StressStretchMullins}a shows the evolution of the soft volume fraction $\nu_s$, during a load/unload/reload cycle, for a representative material with parameters tabulated in Table \ref{tab:MaterialParameters}. 
\begin{table}[htb]
\centering
\begin{tabular}{cc}
\hline
Parameter & Value \\
\hline
$G_0$ (kPa) & $199.26$ \\
$K (= 10^3 G_0)$ (MPa) & $199.26$ \\ 
$A$  & $0.60$ \\
$\nu_{s0} $ & $0.60$ \\
$\nu_{ss} $ & $0.95$ \\
$\lambda_L$ & $2.04$ \\
\hline
\end{tabular}
\caption{Material parameters for a representative rubber-like material.}
\label{tab:MaterialParameters}
\end{table}
During the first loading cycle, $\nu_s$ evolves according to \eqref{eq:EvolutionSoftDomain} and \eqref{eq:LamDotMax}, such that changes occur only when the stretch exceeds the prior maximum value. Upon unloading, $\nu_s$ remains constant as expected. During the second loading, $\nu_s$ remains unchanged until the stretch exceeds the prior maximum value, after which, it evolves again according to \eqref{eq:EvolutionSoftDomain} and \eqref{eq:LamDotMax}. The same behavior is observed in the third loading cycle. Laslty, Figure \ref{fig:StressStretchMullins}b represents the corresponding Cauchy stress vs. stretch response with the Cauchy stress given in \eqref{eq:CauchyStress}.

\subsection{Non-local constitutive model: Helmholtz equation-type and Laplacian non-local formulations}\label{sec:Summary}

\subsubsection{Helmholtz equation-type formulation}
\paragraph{Free energy. Cauchy stress. Evolution equations. Governing equations}

Since our goal is to extend the model of \citet{qi2004constitutive}, we adopt the same\footnote{As discussed later, the non-local formulation adopts the non-local soft volume fraction $\nu_s^{\text{nonlocal}}$ in place of $\nu_s$.} free energy density per unit reference volume as given in \eqref{eqn:FreeEnergy}, which we reproduce here for completeness
\begin{multline*}
\hat{\psi}_\mat(\bfC,\nu_s^{\text{nonlocal}} )  = 
 \nu_s^{\text{nonlocal}} G_0 {\lambda_L}^{2}\left[ \left( \frac{\Lambda}{\lambda_L} \right) \beta + \ln\left( \frac{\beta}{\sinh\beta}   \right) - \left( \frac{1}{\lambda_L}\right) \beta_0 - \ln\left(\frac{\beta_0}{\sinh\beta_0}\right)\right] \\
   + \frac{1}{2} K ({\ln J})^{2} \,.
\end{multline*}

Accordingly, the Cauchy stress is identical to that presented in \eqref{eq:CauchyStress} and is also reproduced here for completeness
\begin{equation*}
	\bfT=J^{-1}\bfF\left(2\dfrac{\partial\hat{\psi}_\mat}{\partial\bfC}\right)
	\bfF^{\trans} = J^{-1}\big[ \nu_s^{\text{nonlocal}} X G_0\left(\frac{\lambda_L}{3\Lambda}\right) \calL^{-1}\left(\frac{\Lambda}{\lambda_L}\right) \left(\bfB_\dis\right)_{0} + K \left(\ln J\right) \id \big]
	\, .
\end{equation*}

We note that in the model developed by \citet{qi2004constitutive}, the soft volume fraction $\nu_s$, evolves according to \eqref{eq:EvolutionSoftDomain}, and is treated as a local internal variable. As a result, its evolution is solved at the material point level, which leads to mesh dependence inherent to the local formulation.
To address this limitation using our first approach, we introduce a Helmholtz-type equation to regularize the soft volume fraction such that 
\begin{equation}\label{eqn:HelmholtzType}
    \tau \dot{\nu_s} = \nu_s - \nu_s^{\text{nonlocal}} + \ell^2 \ \divx (\grad \nu_s^{\text{nonlocal}}) \,, \quad \nu_s^{\text{nonlocal}} (\bfx_\mat, t = 0) = \nu_{s0}\,, 
\end{equation}
along with boundary conditions
\begin{equation}\label{eqn:EvolutionSoftDomain_NonLocal_HelmholtzType_BC}   
  \left.
  \begin{aligned}
 \nu_s^{\text{nonlocal}} = 0 \qquad  \qquad &\text{on} \quad \Sdnonlocal\,,\\
       \grad \nu_s^{\text{nonlocal}} \cdot \bfn_\mat = 0  \qquad &\text{on} \quad \Sgradnunonlocal\,,
  \end{aligned}  
  \right\}
\end{equation}
where $\ell>0$ represents an intrinsic length scale that regularizes the spatial evolution of $\nu_s^{\text{nonlocal}}$, and $\tau>0$ is a parameter for viscous regularization. $\Sdnonlocal$ and $\Sgradnunonlocal$ represent complementary subsurfaces of the boundary $\partial\Br$ of the body $\Br$, such that $\partial\Br = \Sdnonlocal \cup \Sgradnunonlocal$ and $\Sdnonlocal \cap \Sgradnunonlocal = \varnothing$,  with an initial condition $\nu_s^{\text{nonlocal}}(\bfx_\mat,0) = \nu_{s0}$ in $\Br$. 

The key distinction introduced here is that the evolution equation in \eqref{eq:EvolutionSoftDomain} is used to update the local internal variable $\nu_s$, while its non-local counterpart, $\nu_s^{\text{nonlocal}}$, is obtained by solving the Helmholtz equation \eqref{eqn:HelmholtzType}.
Accordingly, $\nu_s^{\text{nonlocal}}$ represent a spatial field governed by the partial differential equation (PDE) in \eqref{eqn:HelmholtzType}.

Lastly, as in the local formulation, the displacement field is governed by the balance of forces and moments, which may be expressed in the \emph{spatial} configuration as in \eqref{eqn:BalanceForcesMoments_Spatial}.

\subsubsection{Laplacian type formulation}

\paragraph{Free energy. Cauchy stress. Evolution equations. Governing equations}

Once again, since our goal is to extend the model of \citet{qi2004constitutive}, we adopt the same\footnote{Similarly here, the non-local formulation adopts the non-local soft volume fraction $\nu_s^{\text{nonlocal}}$ in place of $\nu_s$.} free energy density per unit reference volume as given in \eqref{eqn:FreeEnergy}, which we reproduce here for completeness
\begin{multline*}
	\hat{\psi}_\mat(\bfC,\nu_s^{\text{nonlocal}} )  = 
	\nu_s^{\text{nonlocal}} G_0 {\lambda_L}^{2}\left[ \left( \frac{\Lambda}{\lambda_L} \right) \beta + \ln\left( \frac{\beta}{\sinh\beta}   \right) - \left( \frac{1}{\lambda_L}\right) \beta_0 - \ln\left(\frac{\beta_0}{\sinh\beta_0}\right)\right] \\
	+ \frac{1}{2} K ({\ln J})^{2} \,.
\end{multline*}

Similarly, the Cauchy stress is identical to that presented in \eqref{eq:CauchyStress} and is also reproduced here for completeness
\begin{equation*}
\bfT=J^{-1}\bfF\left(2\dfrac{\partial\hat{\psi}_\mat}{\partial\bfC}\right)
\bfF^{\trans} = J^{-1}\big[ \nu_s^{\text{nonlocal}} X G_0\left(\frac{\lambda_L}{3\Lambda}\right) \calL^{-1}\left(\frac{\Lambda}{\lambda_L}\right) \left(\bfB_\dis\right)_{0} + K \left(\ln J\right) \id \big]
  \, .
\end{equation*}

As noted earlier, the soft volume fraction $\nu_s$ in the model developed by \citet{qi2004constitutive} evolves according to \eqref{eq:EvolutionSoftDomain} and is treated as a local internal variable. As a result, its evolution is solved at the material point level, which leads to mesh dependence inherent to the local formulation.

Accordingly, in our second approach, motivated by its phenomenological nature and the earlier work of Aifantis and co-workers \citep{aifantis1987physics,muhlhaus1991variational}, we extend the evolution of the soft volume fraction in a straightforward manner by augmenting the local evolution law with a Laplacian term such that 
\begin{multline}\label{eq:EvolutionSoftDomain_NonLocal_Spatial}
\dot{\nu_s}^{\text{nonlocal}} = A (\nu_{ss}-\nu_s^{\text{nonlocal}}) \frac{\lambda_{L} - 1}{(\lambda_{L} - \Lambda^\text{max})^2} \dot{\Lambda}^\text{max} \\ + \frac{\ell^2}{\tau} \divx (\grad \nu_s^{\text{nonlocal}})\,, \quad \nu_s^{\text{nonlocal}} (\bfx_\mat, t = 0) = \nu_{s0}\,,
\end{multline}
with
\begin{equation}\label{eq:LamDotMax_NonLocal}
  \dot{\Lambda}^\text{max} =\begin{cases}
    \dot{\Lambda}, &  \Lambda = \Lambda^\text{max}\,,\\
      0 &  \Lambda < \Lambda^\text{max}\,,
  \end{cases} 
\end{equation}
along with boundary conditions
\begin{equation}\label{eqn:EvolutionSoftDomain_NonLocal_Spatial_BC}   
  \left.
  \begin{aligned}
 \nu_s^{\text{nonlocal}} = 0 \qquad  \qquad &\text{on} \quad \Sdnonlocal\,,\\
       \grad \nu_s^{\text{nonlocal}} \cdot \bfn_\mat = 0  \qquad &\text{on} \quad \Sgradnunonlocal\,,
  \end{aligned}  
  \right\}
\end{equation}
where $\ell>0$ represents an intrinsic length scale that regularizes the spatial evolution of $\nu_s^{\text{nonlocal}}$, and $\tau>0$ is a parameter for viscous regularization.  $\Sdnonlocal$ and $\Sgradnunonlocal$ represent complementary subsurfaces of the boundary $\partial\Br$ of the body $\Br$, such that $\partial\Br = \Sdnonlocal \cup \Sgradnunonlocal$ and $\Sdnonlocal \cap \Sgradnunonlocal = \varnothing$,  with an initial condition $\nu_s^{\text{nonlocal}}(\bfx_\mat,0) = \nu_{s0}$ in $\Br$. With this extension, $\nu_s^{\text{nonlocal}}$ is no longer treated as a local internal variable, but instead as a spatial field governed by the PDE in \eqref{eq:EvolutionSoftDomain_NonLocal_Spatial}.

Lastly, as in the local formulation, the displacement field is governed by the balance of forces and moments, which may be expressed in the \emph{spatial} configuration as in \eqref{eqn:BalanceForcesMoments_Spatial}.

\section{Finite element implementation}\label{sec:FE_Implementation}

In this section, we explore the similarities between the heat equation and the non-local soft volume fraction evolution by examining an Abaqus finite element implementation based on our recent work \citep{ALKHOURY2026}.

\subsection{Abaqus implementation: Analogy to the heat equation}

As per the Abaqus documentation \citep{AbqStan}, the heat equation (energy balance) is given by

\begin{equation}\label{eqn:HeatEquation_Abaqus}
\int_V \rho \dot{U}\,dV = \int_S q\,dS + \int_V r\,dV\,,
\end{equation}
where $V$ is the volume of solid material with surface area $S$, $\rho$ is the density of the material, $\dot{U}$ is the material time rate of the internal thermal energy, $q$ is the heat flux per unit area flowing into the body, and $r$ is the heat supplied externally into the body per unit volume.
Using the divergence theorem, along with Fourier's law and the important relation $\dot{U} \Def C \dot{\theta}$, with $C$ representing the specific heat measured in energy per unit mass per temperature for a fixed deformation, and $\dot{\theta}$ the rate of change of temperature, the heat equation \eqref{eqn:HeatEquation_Abaqus} may be recast into its strong form 
\begin{equation} \label{eq:HeatEquation_StrongForm}
    \rho C \dot{\theta} = \divx (\bfK \grad \theta) + r\,,
\end{equation}
where $\bfK$ is the thermal conductivity tensor.

We start by recasting \eqref{eqn:HelmholtzType} such that \begin{equation}
    \tau \dot{\nu_s}^{\text{nonlocal}} = \ell^2 \ \divx (\grad \nu_s^{\text{nonlocal}}) + \underbrace{\nu_s - \nu_s^{\text{nonlocal}}}_r  \,, 
    \label{eqn:DamageEvolution_Spatial_RECAST_Helmholtz}
\end{equation}
and use it in Table \ref{tab:comparison} to show its resemblance to the heat equation using a term-by-term comparison.

We similarly recast \eqref{eq:EvolutionSoftDomain_NonLocal_Spatial} such that
\begin{equation}\label{eqn:DamageEvolution_Spatial_RECAST}
    \dot{\nu_s}^{\text{nonlocal}}  =  \frac{\ell^2}{\tau} \divx (\grad \nu_s^{\text{nonlocal}}) + \underbrace{A (\nu_{ss}-\nu_s^{\text{nonlocal}}) \frac{\lambda_{L} - 1}{(\lambda_{L} - \Lambda^\text{max})^2} \dot{\Lambda}^\text{max}}_r\,,
\end{equation}
and use it in Table \ref{tab:comparison} to show its resemblance to the heat equation using a term-by-term comparison.
\begin{table}[h]
    \centering
    \renewcommand{\arraystretch}{1.5} 
    \begin{tabular}{c|c|c|c}
       \textbf{Equation}  & \textbf{Transient term} & \textbf{Conduction term} & \textbf{Source term} \\ \hline
        \textbf{ \eqref{eq:HeatEquation_StrongForm}} & $\rho C \dot{\theta}$ & $\divx (\bfK \grad \theta)$ & $r$ \\ 
        \textbf{ \eqref{eqn:DamageEvolution_Spatial_RECAST_Helmholtz}} & $\tau \dot{\nu_s}^{\text{nonlocal}}$ & $ \ell^2 \divx (\grad \nu_s^{\text{nonlocal}})$ & $\nu_s - \nu_s^{\text{nonlocal}} $ \\
        \textbf{ \eqref{eqn:DamageEvolution_Spatial_RECAST}} & $\dot{\nu_s}^{\text{nonlocal}}$ & $ \frac{\ell^2}{\tau} \divx (\grad \nu_s^{\text{nonlocal}})$ & $A (\nu_{ss}-\nu_s^{\text{nonlocal}}) \frac{\lambda_{L} - 1}{(\lambda_{L} - \Lambda^\text{max})^2} \dot{\Lambda}^\text{max}$
    \end{tabular}
    \caption{Term-by-term comparison of the heat equation \eqref{eq:HeatEquation_StrongForm} with the non-local soft volume evolution equations: (i) Helmholtz-type formulation \eqref{eqn:DamageEvolution_Spatial_RECAST_Helmholtz} and (ii) Laplacian-type formulation \eqref{eqn:DamageEvolution_Spatial_RECAST}.}
    \label{tab:comparison}
\end{table}

In order to use the heat equation provided by Abaqus through the user subroutine UMAT, one needs to modify each of the terms in Table \ref{tab:comparison} as follows, with the caveat that the non-local soft volume fraction ``$\nu_s^{\text{nonlocal}}$'' is represented by temperature ``$\theta$'':

\begin{enumerate}
\item Helmholtz equation-type formulation: \begin{enumerate}
    \item Starting with the transient term, one needs to impose the equality $\rho C  = \tau$, which can be easily achieved through the input file by setting $\rho = \tau$ and $C=1$, without modifying the user subroutine. 
    \item Next, one needs to set the thermal conductivity tensor $\bfK$ in Fourier's law to the identity tensor $\bfI$ (or equivalently to a scalar value of $k=1$) through the input file. The conduction term is then multiplied by a pre-factor $\ell^2$, which is introduced in the user subroutine ``UMATHT'' as the variable ``AUX,'' as shown in the code\footnote{We refer to the original code provided by Abaqus documentation as ``original code'' and the modifications done in this work as ``modified code.'' } below:
\begin{itemize}
    \item original code: \begin{lstlisting}[language=Fortran, firstnumber=20]
    DO I=1, NTGRD
        FLUX(I) = -COND*DTEMDX(I)
        DFDG(I,I) = -COND
    END DO
    \end{lstlisting}
    \item modified code: \begin{lstlisting}[language=Fortran, firstnumber=20]
    AUX = (lc**two)
    DO I=1, NTGRD
        FLUX(I) = -COND*DTEMDX(I)
        DFDG(I,I) = -COND
    END DO
    FLUX = FLUX * AUX
    DFDG = AUX * DFDG

      \end{lstlisting}
\end{itemize}
\item Lastly, the source term needs to be $r = \nu_s - \nu_s^{\text{nonlocal}}$ , which can be directly achieved through the ``RPL'' functionality in the user subroutine ``UMAT'' with the details provided in the code. 

And, since DRPLDT ($\frac{\partial r}{\partial \theta} = \frac{\partial r}{\partial \nu_s^{\text{nonlocal}}}$), the variation of RPL with respect to temperature (soft volume fraction) is needed for the Newton solver to ensure convergence in coupled ``temperature'' - displacement analyses, we also introduce 
\begin{equation}\label{eqn:drpldt}
  \frac{\partial r}{\partial \nu_s^{\text{nonlocal}}} = - 1 \,.
\end{equation}
\end{enumerate}
    \item Laplacian type formulation: \begin{enumerate}
    \item Starting with the transient term, one needs to impose the equality $\rho C  = 1$, which can be easily achieved through the input file without modifying the user subroutine. 
    \item Next, one needs to set the thermal conductivity tensor $\bfK$ in Fourier's law to the identity tensor $\bfI$ (or equivalently to a scalar value of $k=1$) through the input file. The conduction term is then multiplied by a pre-factor $\frac{\ell^2}{\tau}$, which is introduced in the user subroutine ``UMATHT'' as the variable ``AUX,'' as shown in the code\footnote{Once again, we refer to the original code provided by Abaqus documentation as ``original code'' and the modifications done in this work as ``modified code.'' } below:
\begin{itemize}
    \item original code: \begin{lstlisting}[language=Fortran, firstnumber=20]
    DO I=1, NTGRD
        FLUX(I) = -COND*DTEMDX(I)
        DFDG(I,I) = -COND
    END DO
    \end{lstlisting}
    \item modified code: \begin{lstlisting}[language=Fortran, firstnumber=20]
    AUX = (lc**two)/tau
    DO I=1, NTGRD
        FLUX(I) = -COND*DTEMDX(I)
        DFDG(I,I) = -COND
    END DO
    FLUX = FLUX * AUX
    DFDG = AUX * DFDG

      \end{lstlisting}
\end{itemize}
\item Lastly, the source term needs to be $r = A (\nu_{ss}-\nu_s^{\text{nonlocal}}) \frac{\lambda_{L} - 1}{(\lambda_{L} - \Lambda^\text{max})^2} \dot{\Lambda}^\text{max}$, which can be directly achieved through the ``RPL'' functionality in the user subroutine ``UMAT'' with the details provided in the code. 

And, since DRPLDT ($\frac{\partial r}{\partial \theta} = \frac{\partial r}{\partial \nu_s^{\text{nonlocal}}}$), the variation of RPL with respect to temperature (soft volume fraction) is needed for the Newton solver to ensure convergence in coupled ``temperature'' - displacement analyses, we also introduce 
\begin{equation}\label{eqn:drpldt}
  \frac{\partial r}{\partial \nu_s^{\text{nonlocal}}} = - A  \frac{\lambda_{L} - 1}{(\lambda_{L} - \Lambda^\text{max})^2} \dot{\Lambda}^\text{max} \,.
\end{equation}
\end{enumerate}
\end{enumerate}

\section{Application to modeling the inhomogeneous cyclic loading in a rubber-like material}\label{sec:Applications}
In this section, we demonstrate the relevance of our non-local model by considering a boundary-value problem and comparing its results with those of the local model by \citet{qi2004constitutive}.

\begin{figure}[H]
	\centering
	\includegraphics[width = .5\textwidth]{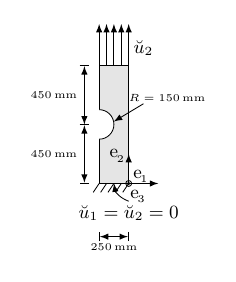}
	\caption{Schematic of the boundary value problem of a rubber-like material subjected to a cyclic loading profile in uniaxial tension.}
	\label{fig:BVP_SENT}
\end{figure}

We consider the inhomogeneous deformation of a notched representative rubber-like material of length $L_0 = 900$\,mm, width $W_0 = 250$\,mm and notch radius $R = 150$\,mm with geometry and boundary conditions shown in Figure \ref{fig:BVP_SENT}, and material parameters provided in Table \ref{tab:MaterialParametersNonLocal}.

\begin{table}[htb]
\centering
\renewcommand{\arraystretch}{1.15}
\begin{tabular}{c|c|p{3.2cm}|p{3.2cm}}
\hline
 & Local & \multicolumn{2}{c}{Non-local} \\
\cline{3-4}
Parameter &  & \centering (i) Helmholtz equation-type formulation
& \centering (ii) Laplacian type formulation \tabularnewline
\hline
$G_0$ (kPa)              & $199.26$ & $199.26$ & $199.26$ \\
$K (=10^3 G_0)$ (MPa)    & $199.26$ & $199.26$ & $199.26$ \\
$A$                      & $0.60$   & $0.60$   & $0.60$   \\
$\nu_{s0}$               & $0.60$   & $0.60$   & $0.60$   \\
$\nu_{ss}$               & $0.95$   & $0.95$   & $0.95$   \\
$\lambda_L$              & $2.04$   & $2.04$   & $2.04$   \\
$\ell$ (mm)              & ---      & $0.2$   & $0.1$   \\
$\tau$ (s)              & ---    & $10^{-2}$   & $1$   \\
\hline
\end{tabular}
\caption{Material parameters for a representative rubber-like material for both the local and non-local formulations.}
\label{tab:MaterialParametersNonLocal}
\end{table}

\begin{figure}[H]
\centering
\includegraphics[width = .5\textwidth]{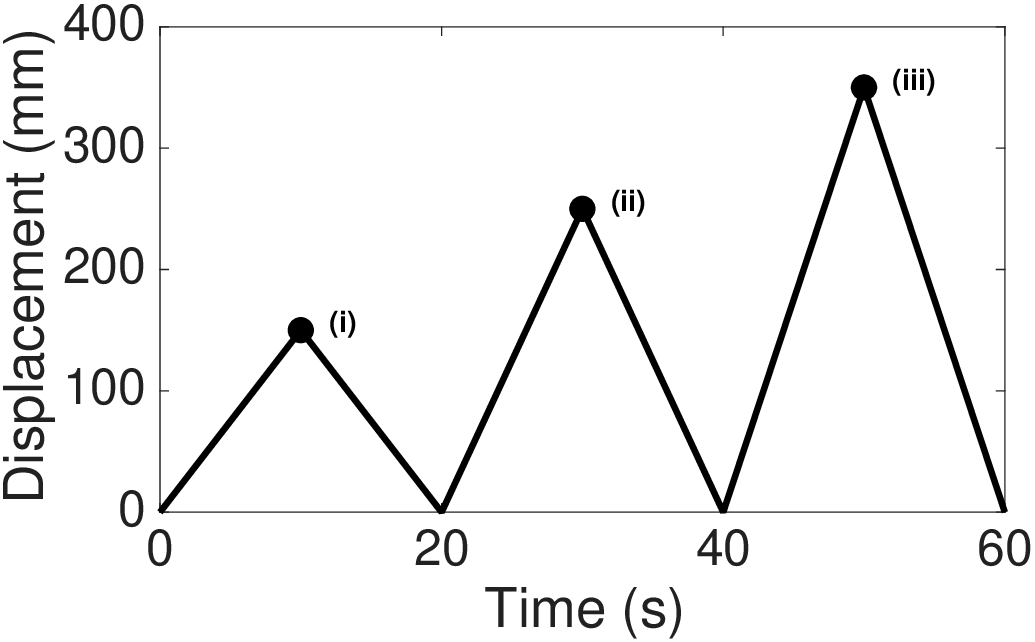}
	\caption{Prescribed cyclic displacement-time loading profile.}
\label{fig:BVP_DisplacementProfile}
\end{figure}

The bottom surface is fixed, and a cyclic displacement profile shown in Figure \ref{fig:BVP_DisplacementProfile} is applied to the top surface, while all other surfaces are traction-free.

\begin{figure}[H]
\centering
\begin{tabular}{ccc}
\includegraphics[width = .25\textwidth]{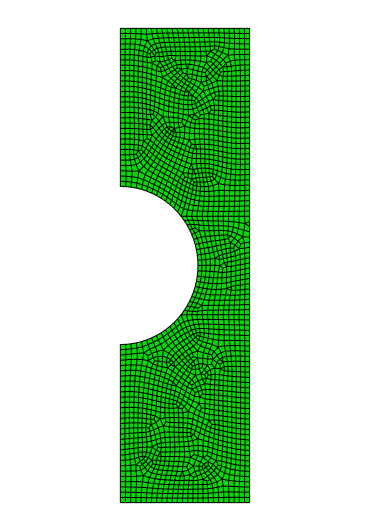}  & \includegraphics[width = .25\textwidth]{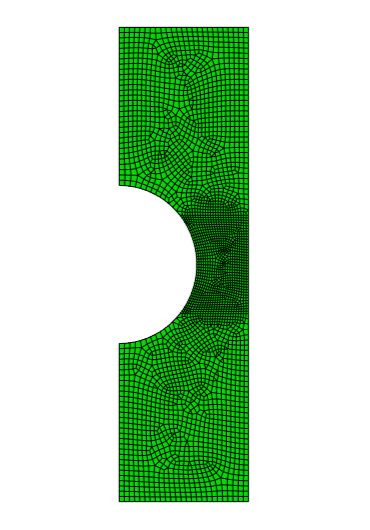} &
\includegraphics[width = .25\textwidth]{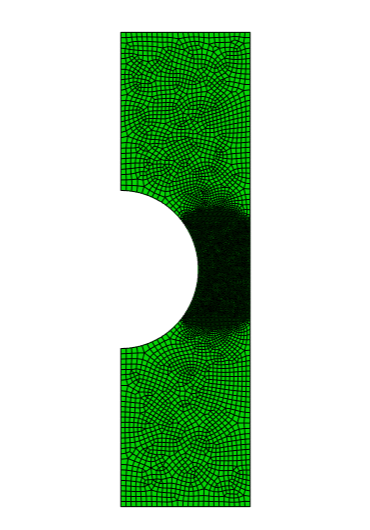}\\
(i) coarse & (ii) medium & (iii) fine
    \end{tabular}
	\caption{Finite element meshes used to assess mesh dependence near the notch: (i) coarse ($l_e = 10$ mm), (ii) medium ($l_e = 5$ mm), and (iii) fine ($l_e = 2.5$ mm), with refinement concentrated in the notch region.}
\label{fig:Mesh}
\end{figure}

In both cases, local and non-local formulations, the identical geometry is discretized using a plane-strain approximation with a combination of 3-node and 4-node linear displacement-temperature elements (CPE3T and CPE4T). To assess mesh dependence, we consider three levels of mesh refinement near the notch, which is expected to act as a stress concentration region and localize the Mullins effect: (i) \emph{coarse}, with average element size of $l_e = 10$\,mm, (ii) \emph{medium}, with average element size of $l_e =  5$\,mm, and (iii) \emph{fine}, with average element size of $l_e =  2.5$\,mm, as shown in Figure \ref{fig:Mesh}.

\begin{figure}[H]
\centering
 Local -- \citet{qi2004constitutive} \\
\begin{tabular}{ccc}
\includegraphics[width = .25\textwidth]{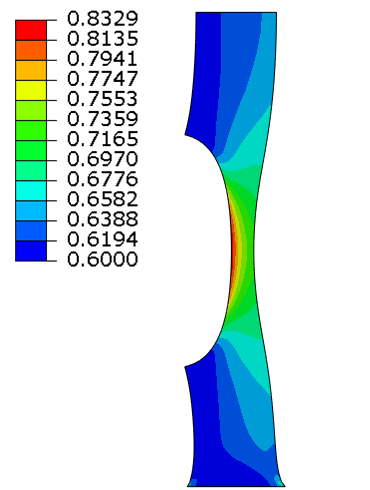}  & \includegraphics[width = .25\textwidth]{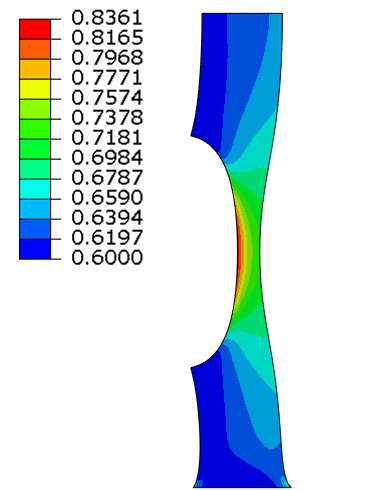} &
\includegraphics[width = .25\textwidth]{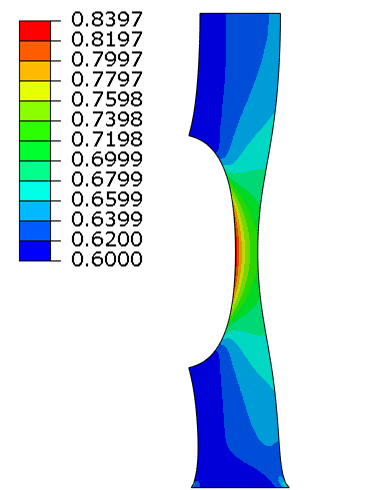}\\
a) coarse & b) medium & c) fine 
\end{tabular}
\\
\centering
Non-local -- (i) Helmholtz equation-type formulation \\
\begin{tabular}{ccc}
\includegraphics[width = .25\textwidth]{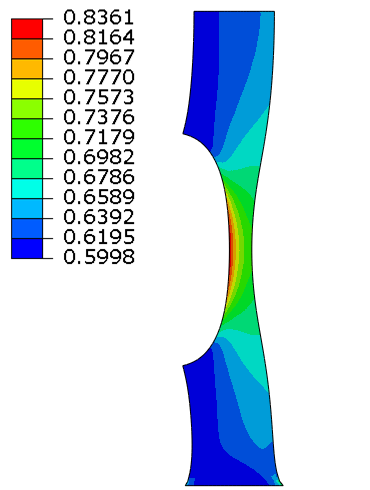}  & \includegraphics[width = .25\textwidth]{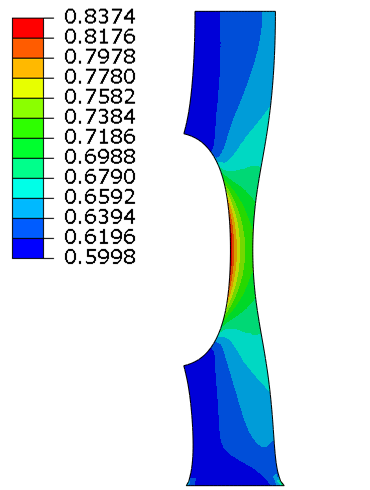} &
\includegraphics[width = .25\textwidth]{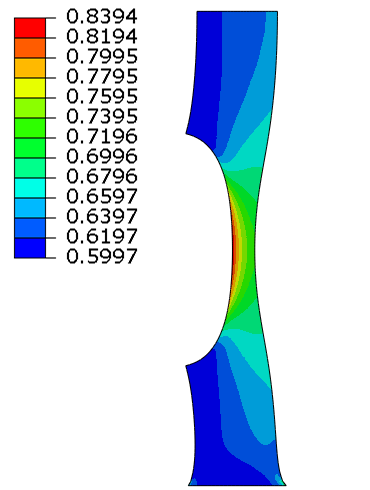}\\
d) coarse & e) medium & f) fine
\end{tabular}
\\
\centering
Non-local -- (ii) Laplacian type formulation\\
\begin{tabular}{ccc}
	\includegraphics[width = .25\textwidth]{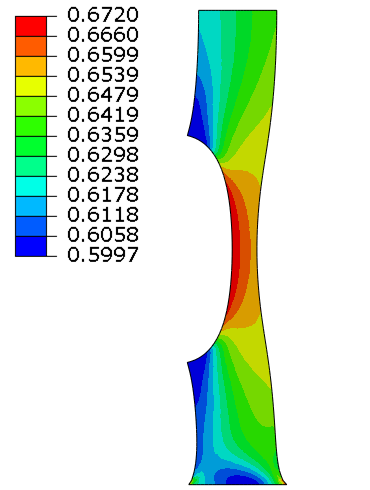}  & \includegraphics[width = .25\textwidth]{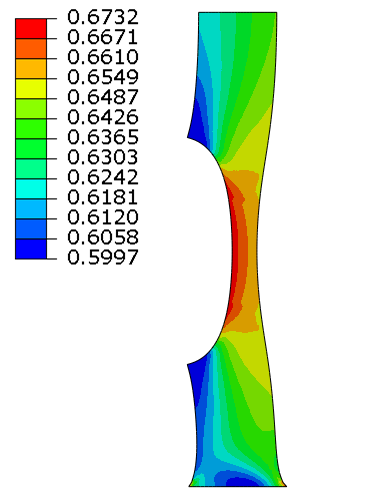} &
	\includegraphics[width = .25\textwidth]{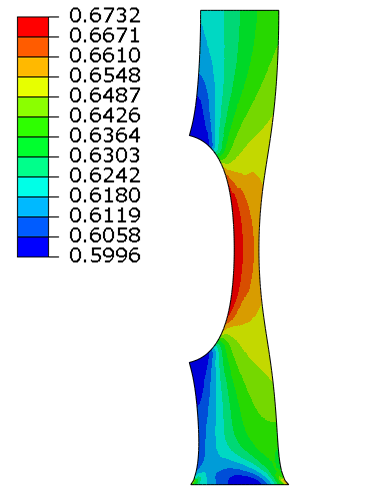}\\
	g) coarse & h) medium & i) fine
\end{tabular}
\caption{Comparison of the soft volume fraction evolution at instant (iii) in Figure \ref{fig:BVP_DisplacementProfile}, for the same notch geometry: (a--c) local model of \citet{qi2004constitutive} using a) coarse ($l_e = 10$ mm), b) medium ($l_e = 5$ mm), and c) fine ($l_e = 2.5$ mm) meshes; (d--f) the non-local Helmholtz equation-type formulation developed in this work using d) coarse ($l_e = 10$ mm), e) medium ($l_e = 5$ mm), and f) fine ($l_e = 2.5$ mm) meshes; and (g--i) the non-local Laplacian type formulation developed in this work using g) coarse ($l_e = 10$ mm), h) medium ($l_e = 5$ mm), and i) fine ($l_e = 2.5$ mm) meshes.}
\label{fig:SoftVolume_Reload2}
\end{figure}

Figure \ref{fig:SoftVolume_Reload2} reports on the corresponding soft volume fraction evolution at instant (iii), for the local model\footnote{We note that the local results were obtained from the non-local (Helmholtz equation-type formulation) simulation results using the local soft volume fraction. A separate implementation of the purely local formulation is provided for completeness, although it is not used here. Any slight discrepancies are attributed to differences in time stepping and solver accuracy between the coupled displacement–temperature and displacement-only analyses.} by \citet{qi2004constitutive} and the two non-local models developed in this work, for various mesh refinements, and the results for all each condition at each instant are tabulated in Table \ref{tab:nus_comparison}. Note that we report the maximum soft volume fraction for both the local and non-local cases at different instants, and the percent difference is defined such that $\calE_r \Def \frac{|\text{fine} - {\text{coarse}|}}{\text{{fine}}} \times 100 \%$\,.

\begin{table}[htb]
\centering
\setlength{\tabcolsep}{5pt}
\renewcommand{\arraystretch}{1.15}
\resizebox{\textwidth}{!}{%
\begin{tabular}{c cccc cccc cccc}
\toprule
& \multicolumn{4}{c}{Local} 
& \multicolumn{4}{c}{Non-local: Helmholtz equation-type formulation} 
& \multicolumn{4}{c}{Non-local: Laplacian type formulation} \\
\cmidrule(lr){2-5} \cmidrule(lr){6-9} \cmidrule(lr){10-13}
Instants 
& \multicolumn{4}{c}{$\nu_s$}
& \multicolumn{4}{c}{$\nu_s^{\text{nonlocal}}$}
& \multicolumn{4}{c}{$\nu_s^{\text{nonlocal}}$} \\
\cmidrule(lr){2-5} \cmidrule(lr){6-9} \cmidrule(lr){10-13}
& coarse & medium & fine & $\mathcal{E}_r$ (\%)
& coarse & medium & fine &  $\mathcal{E}_r$ (\%)
& coarse & medium & fine & $\mathcal{E}_r$ (\%) \\
\midrule
(i) 
& 0.6884  & 0.6903  & 0.6913 & \textbf{0.42}
& 0.6903 & 0.6913  & 0.6917 & \textbf{0.20}
& 0.6549 & 0.6547 & 0.6547  & \textbf{0.03} \\

(ii) 
& 0.7666 & 0.7697 & 0.7714 & \textbf{0.63}
& 0.7696 & 0.7712  & 0.7719 & \textbf{0.30}
& 0.6685 & 0.6680 & 0.6696 & \textbf{0.17} \\

(iii) 
& 0.8329 & 0.8361 & 0.8397 & \textbf{0.82}
& 0.8361 & 0.8374  & 0.8394 & \textbf{0.40}
& 0.6720 & 0.6732 & 0.6732 & \textbf{0.18} \\
\bottomrule
\end{tabular}%
}
\caption{Comparison of the maximum local soft volume fraction $\nu_s$, and its non-local counterpart $\nu_s^{\text{nonlocal}}$, obtained with the coarse, medium, and fine meshes for the local and non-local approaches at different instants. \\ Note that the Helmholtz equation-type formulation effectively recovers the local response, while reducing mesh dependence through its regularizing effect. In contrast, the Laplacian type formulation introduces a stronger regularization, resulting in a more spatially distributed and less localized soft volume fraction field.} 
\label{tab:nus_comparison}
\end{table}

It can be observed that $\calE_r$ is consistently higher for the local formulation at all instants, whereas it is lower for both non-local formulations, indicating improved mesh-independent behavior. The Helmholtz equation-type formulation reduces the error by approximately a factor of two, whereas the Laplacian type formulation exhibits a more pronounced reduction, with differences exceeding an order of magnitude in some cases.
Moreover, differences in the spatial distribution of the soft volume fraction can be observed across the formulations. The Helmholtz equation-type formulation yields nearly identical soft volume fraction values compared to the local formulation, but with reduced mesh dependence due to the regularizing effect of the formulation. In contrast, the Laplacian type formulation produces a more spatially distributed and smoother field, resulting in reduced localization as a consequence of the stronger regularization.
Lastly, although the differences in the local $\nu_s$ and non-local $\nu_s^{\text{nonlocal}}$ soft volume fractions appear small, it is important to emphasize that the soft volume fraction evolves over a relatively narrow range (e.g., from $\nu_{s0} \approx 0.65$ to $\nu_{ss} \approx 0.95$ in the present example, and in some situations over an even smaller interval). Consequently, and especially since filled elastomers are typically subjected to a very large number of loading cycles during service life, even minor numerical differences can accumulate and lead to significant deviations in the predicted long-term response, and therefore should not be ignored.

\begin{Remark}
We note that the Helmholtz equation-type formulation effectively recovers the local response, while reducing mesh dependence through its regularizing effect. In contrast, the Laplacian type formulation introduces a stronger regularization, resulting in a more spatially distributed and less localized soft volume fraction field, as observed in the results in Figure \ref{fig:SoftVolume_Reload2} and Table \ref{tab:nus_comparison}.
\end{Remark}
\begin{Remark}
We also note that the apparent increase in the non-local soft volume fraction for the Helmholtz equation-type formulation relative to the local formulation is a post-processing artifact (cf. Figure \ref{fig:SoftVolume_Reload2}a and d, Figure \ref{fig:SoftVolume_Reload2}b and e, and Figure \ref{fig:SoftVolume_Reload2}c and f). This arises from displaying the results using a solution-depedent state variable (SDV) for consistency, as Abaqus employs different interpolation procedures for temperature (NT11) and SDVs.
\end{Remark}

\section{Conclusion}\label{sec:Conclusion}

In this work, we extend an established local constitutive model for the Mullins effect in filled elastomers to a non-local setting to address the mesh dependence inherent to the local approach. This was accomplished using two approaches: (i) a Helmholtz-type equation governing a non-local soft volume fraction, and (ii) a Laplacian term introduced directly into the soft volume fraction local evolution law. In both formulations, an additional governing partial differential equation arises and is solved numerically in Abaqus using the analogy with the heat equation. The two approaches yield different results, leaving the choice between them to be guided by experimental findings. The details of the implementation using the user material subroutines UMAT and UMATHT are provided and were further used to study a boundary value problem to demonstrate the relevance of our non-local model. The Abaqus user material subroutines developed in this work, which include the constitutive details, along with the input files, are provided as supplemental materials to this paper. 

Although the present study focuses on the quasi-static mechanical response of filled elastomers, many rubber-like materials exhibit pronounced rate-dependent behavior due to viscoelastic effects. Accordingly, extending the present non-local formulation to account for viscoelasticity represents a natural direction for future work, particularly in light of recent studies, including ours, suggesting that the Mullins effect and viscoelasticity are coupled \citep{alkhoury2024experiments,lamont2026physically}.

\newpage

\section*{CRediT authorship contribution statement}
\textbf{Keven Alkhoury}: Conceptualization, Methodology, Software, Formal analysis, Investigation, Writing - Original Draft.

\section*{Acknowledgments}
The author acknowledges computational hardware and software support from the Hibbitt Engineering Fellowship at Brown University. The author thanks Shawn A. Chester of the New Jersey Institute of Technology (NJIT) for fruitful discussions.

\section*{Code availability}
The code developed and used in this study will be made available upon publication.

\newpage

\bibliographystyle{abbrvnat}
\bibliography{References}

\end{document}